\documentclass[english,usenatbib,onecolumn]{mnras}
\usepackage[T1]{fontenc}
\usepackage[latin9]{inputenc}
\usepackage{geometry}
\usepackage{amsmath}
\usepackage{amssymb}
\usepackage{graphicx}
\usepackage{esint}

\makeatletter

\providecommand{\tabularnewline}{\\}

\numberwithin{equation}{section}

\usepackage{lineno}

\makeatother

\usepackage{babel}
\begin{document}

\title{Large Covariance Matrices: Smooth Models from the 2-Point Correlation
Function}

\author[O'Connell et al.]{Ross O'Connell$^1$, Daniel Eisenstein$^2$, Mariana Vargas$^3$, Shirley Ho$^1$,
\newauthor  Nikhil
Padmanabhan$^4$
\\
$^{1}$McWilliams Center for Cosmology, Carnegie Mellon University, 5000 Forbes Ave, Pittsburgh, PA 15213, USA\\
$^{2}$Harvard-Smithsonian Center for Astrophysics, 60 Garden St., Cambridge, MA 02138, USA\\
$^{3}$Instituto de Física, Universidad Nacional Autónoma de México, Apdo. Postal 20-364, México\\
$^{4}$Dept. of Physics, Yale University, New Haven, CT, 06511, USA
}

\maketitle
\begin{abstract}
We introduce a new method for estimating the covariance matrix for
the galaxy correlation function in surveys of large-scale structure.
Our method combines simple theoretical results with a realistic characterization
of the survey to dramatically reduce noise in the covariance matrix.
For example, with an investment of only $\approx1,000$ CPU hours
we can produce a model covariance matrix with noise levels that would
otherwise require $\sim35,000$ mocks. Non-Gaussian contributions
to the model are calibrated against mock catalogs, after which the
model covariance is found to be in impressive agreement with the mock
covariance matrix. Since calibration of this method requires fewer
mocks than brute force approaches, we believe that it could dramatically
reduce the number of mocks required to analyse future surveys.
\end{abstract}

\section{Introduction}

Covariance matrix estimation is a fundamental challenge in precision
cosmology. In the standard approach many artificial or ``mock''
catalogs are created that mimic the properties of the cosmological
data set, the analysis is performed on each mock catalog, and a sample
covariance computed using those results. From a certain point of view
this approach is very simple, since the mock catalogs are statistically
independent from one another. The potential stumbling block is that
the sample covariance provides a noisy estimate of the true covariance,
and covariance matrix noise can degrade the final measurement. Reducing
noise in the covariance matrix to acceptable levels generally requires
a large number of mock catalogs, and covariance matrix estimation
can consume the majority of computational resources for current
analyses. Moreover, there is a clear trade-off between the number
of mock catalogs required and the accuracy of any individual mock.
We aim to improve this situation with a new method for covariance
matrix estimation that combines simple theoretical results with a
realistic characterization of the survey, and that can be calibrated
against a relatively small number of mocks. We will demonstrate that
this approach produces a covariance matrix suitable for analysis of
baryon acoustic oscillations (BAO) in modern surveys.

The consequences of covariance matrix noise have been a subject of
recent interest \citep{Taylor2013,Dodelson:2013uaa,Percival:2013sga}.
It is generally true that when cosmological parameters are determined
from an observable estimated in $N_{\mathrm{bins}}$ (for example
a correlation function or power spectrum), and the corresponding covariance
matrix is estimated using $N_{\mathrm{mocks}}$ mock catalogs, then
there is a fractional increase in the uncertainty in the cosmological
parameters, relative to an ideal measurement, of $\mathcal{O}\left(1/\left(N_{\mathrm{mocks}}-N_{\mathrm{bins}}\right)\right)$.
In other words, having $N_{\mathrm{mocks}}\sim10\times N_{\mathrm{bins}}$
(rather than much larger) has an impact on the final parameter estimates
comparable to reducing the volume of a survey by $\sim10\%$. When
the covariance matrix is estimated using a sample covariance of mock
catalogs, there is thus an incentive to make $N_{\mathrm{mocks}}-N_{\mathrm{bins}}$
as large as possible.

While some current surveys have achieved $N_{\mathrm{mocks}}\sim100\times N_{\mathrm{bins}}$,
we are concerned that this may not be achievable in future surveys.
First, we expect that many future analyses will attempt to utilize
more accurate mocks, for example incorporating realistic light cones,
and that this will raise the computational cost of a single mock.
In addition, many future surveys (e.g. Euclid \citep{EuclidRedBook},
DESI \citep{Levi:2013gra}, WFIRST-AFTA \citep{WFIRST}) have proposed
to perform tomographic analyses, rather than analysing a single redshift
range; where current analyses might estimate the correlation function
using 40 bins, a tomographic analysis using 10 redshift slices will
necessarily use 400 bins. We emphasize that while available computing
resources may increase at a pace that accommodates increasing mock
requirements, there will always be an opportunity cost to producing
large numbers of mocks -- given fixed computing resources, reducing
the number of mocks required by a factor of 100 means that the amount
of computing time that can be spent on each mock increases by a factor
of 100.

It is possible to address the problem by reducing $N_{\mathrm{bins}}$,
rather than increasing $N_{\mathrm{mocks}}$. For example, current
BAO analyses typically compress the angular dependence of the correlation
function into two multipole moments or a small number of angular wedges,
rather than retaining the full angular dependence of the correlation
function. Proposals have also been made for an analogous compression
of redshift-dependence for tomographic analyses \citep{Zhu:2014ica}.
While these approaches provide an intriguing complement to the method
we will describe, we point out that there may be tension between the
optimal modes for compressing the signal, and the characteristic modes
associated with systematic effects. For example, reducing the angular
dependence of the redshift space galaxy correlation function to monopole
and quadrupole modes provides highly efficient compression of the
underlying correlation function, but systematic effects tend to depend
on line-of-sight or transverse separations between points and thus
compress poorly into the monopole and quadrupole modes.

Motivated by these concerns, we propose a new method for covariance
matrix estimation that accommodates large $N_{\mathrm{bins}}$ while
requiring only a modest number of $N_{\mathrm{mocks}}$, and results
in a dramatic reduction in noise in the covariance matrix relative
to the sample covariance. In Section \ref{sec:Theory} we provide
a simple analytic expression for the covariance matrix of a galaxy
correlation function. This extends the results of Bernstein \citep{Bernstein1994}
to incorporate position-dependent number densities and weights. In
Section \ref{sec:Integrate} we numerically integrate this expression
over a realistic survey geometry, assuming that the underlying galaxy
field is Gaussian. We find reasonable agreement between our results
and a sample covariance computed from 1,000 quick particle mesh (QPM,
\citep{White:2013psd}) mocks produced using the same survey geometry.
In Section \ref{sec:NGModel} we show that increasing the level of
shot noise in the Gaussian model covariance matrix brings it closer
to the full, non-Gaussian covariance matrix. After fitting the shot
noise increase using the QPM mocks, we find exceptional agreement
between our method and the QPM sample covariance.

\section{$n$-Point Functions and the Covariance Matrix}
\label{sec:Theory}

The basic observation that motivates this work is that the covariance
matrix of an $n$-point function can be written as a combination of
2$n-$ and lower-point functions. As a simple example, consider the
correlation function $\xi_{ij}$ of a density field $\delta_{i}$:
\begin{equation}
\xi_{ij}=\delta_{i}\delta_{j}\,.
\end{equation}
The covariance between two observations of the correlation function
is simply 
\begin{eqnarray}
\mathrm{cov}\left(\xi_{ij},\xi_{k\ell}\right) & = & \left\langle \xi_{ij}\xi_{kl}\right\rangle -\left\langle \xi_{ij}\right\rangle \left\langle \xi_{k\ell}\right\rangle \\
 & = & \left\langle \delta_{i}\delta_{j}\delta_{k}\delta_{\ell}\right\rangle -\left\langle \delta_{i}\delta_{j}\right\rangle \left\langle \delta_{k}\delta_{\ell}\right\rangle \\
 & = & \left\langle \xi_{ijk\ell}\right\rangle +\left\langle \xi_{ik}\right\rangle \left\langle \xi_{j\ell}\right\rangle +\left\langle \xi_{i\ell}\right\rangle \left\langle \xi_{jk}\right\rangle ,\label{eq:BasicCov}
\end{eqnarray}
where we have used $\xi_{ijk\ell}$ to denote the connected 4-point
function. For fields that are Gaussian or nearly Gaussian, the value
of expressions like (\ref{eq:BasicCov}) should be clear: they allow
us to convert an estimate of the 2-point correlation function into
an estimate of that correlation function's covariance matrix.

In the following we will demonstrate how to extend this simple result
to correlation functions estimated in bins $\Theta_{ij}^{a}$, how
to modify it for a Poisson-sampled density field, and how to incorporate
the inhomogeneous number density and non-uniform weighting that occur
in realistic surveys. A more comprehensive derivation of these results,
including finite-volume corrections (but without weights and assuming
homogeneity), can be found in \citep{Bernstein1994}. An earlier application
of these ideas, with less attention to reproducing the detailed survey
geometry, can be found in \citep{Xu2013_DR7}.

\subsection{Poisson Sampling}

Suppose that we have divided the volume of a galaxy survey into many
cells, each small enough that it contains at most one galaxy. The
overdensity in cell $i$ is then 
\begin{equation}
\delta_{i}=\frac{b_{i}}{n_{i}}-1\,,
\end{equation}
where $b_{i}=1$ if the cell contains a galaxy and $b_{i}=0$ otherwise,
and $n_{i}$ is the expected number of galaxies in the cell (note
that this may vary from cell to cell). Since the choice of cell size
is arbitrary we can choose $n_{i}\ll1$, and in this limit the overdensity
satisfies a useful identity:
\begin{equation}
\delta_{i}^{2}\approx\frac{1}{n_{i}}\left(1+\delta_{i}\right).\label{eq:Contraction}
\end{equation}
This identity encodes our intuitive understanding of Poisson sampling,
which is that at sufficiently short separations $\left(\delta_{i}\delta_{j}\to\ \delta_{i}^{2}\right)$
the correlation function is replaced by shot noise ($\sim1/n_{i}$).

\begin{figure}
\includegraphics{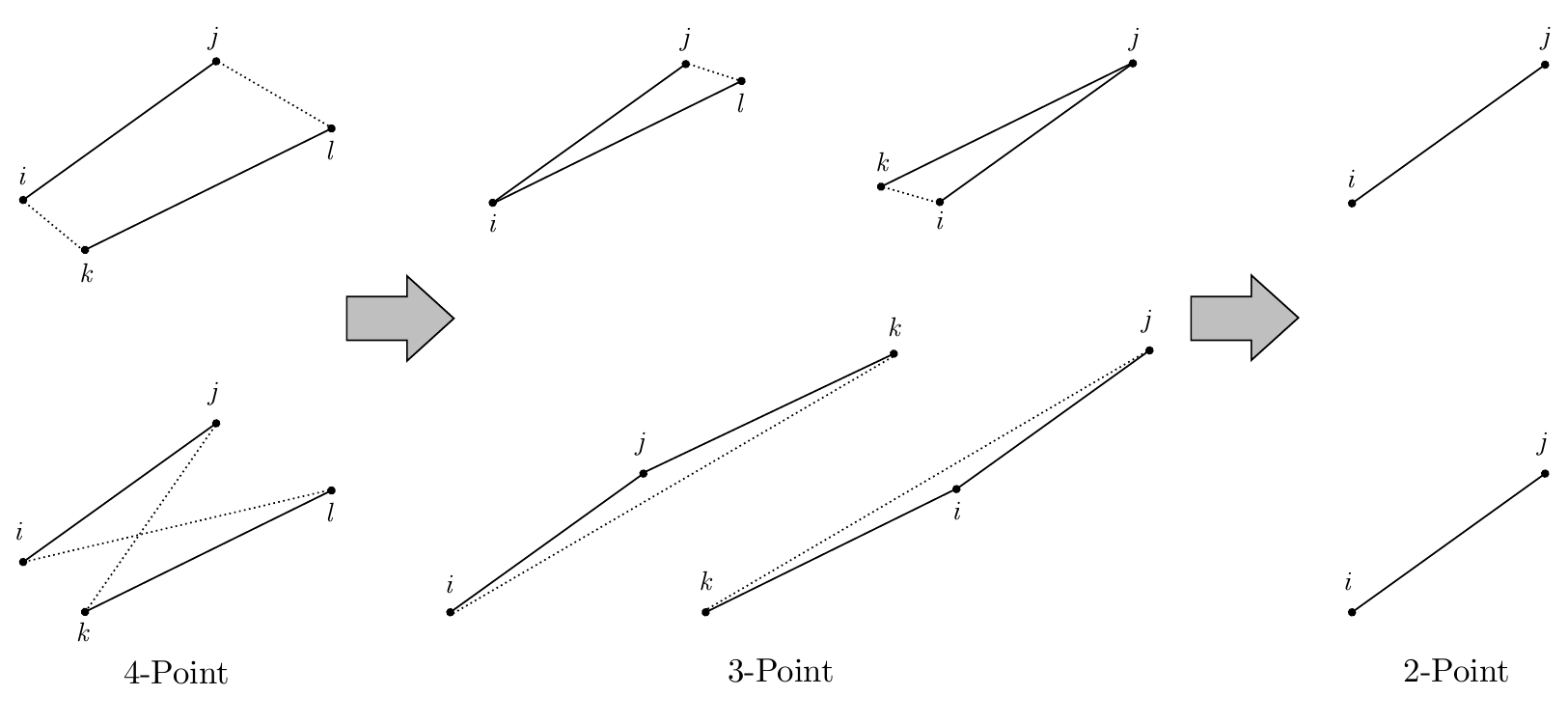}

\caption{\label{fig:Contractions}Configurations of points that contribute
to the covariance matrix integrals. Starting from the two 4-point
configurations that contribute to $C_{4}$, single contractions generate
the four possible 3-point configurations of $C_{3}$, and double contractions
generate the 2-point configuration of $C_{2}$. Solid lines indicate
pairs of points that are governed by a binning function, $\Theta^{a}\left(\vec{r}_{i}-\vec{r}_{j}\right)$,
while dotted lines indicate pairs of points that are governed by a
factor of the correlation function, $\xi\left(\vec{r}_{i}-\vec{r}_{k}\right)$.}
\end{figure}

We estimate the correlation function as 
\begin{eqnarray}
\hat{\xi}^{a} & = & \left(RR^{a}\right)^{-1}\sum_{i\neq j}\Theta_{ij}^{a}n_{i}n_{j}\delta_{i}\delta_{j}\,,\label{eq:xi_estimator}\\
RR^{a} & \equiv & \sum_{i\neq j}\Theta_{ij}^{a}n_{i}n_{j}\,.\label{eq:rr_def}
\end{eqnarray}
The binning matrices $\Theta_{ij}^{a}$ are one for pairs of points
$i,j$ that fall in bin $a$, and zero otherwise. We assume that the
bins require finite separation, so that both sums run over \emph{distinct}
points $i,j$. Note that $RR^{a}$ coincides with the usual notion
of RR pair counts. The covariance of $\hat{\xi}^{a}$ is then 
\begin{equation}
\left\langle \hat{\xi}^{a}\hat{\xi}^{b}\right\rangle -\left\langle \hat{\xi}^{a}\right\rangle \left\langle \hat{\xi}^{b}\right\rangle =\left(RR^{a}RR^{b}\right)^{-1}\sum_{i\neq j,k\neq\ell}\Theta_{ij}^{a}\Theta_{k\ell}^{b}n_{i}n_{j}n_{k}n_{\ell}\left[\left\langle \delta_{i}\delta_{j}\delta_{k}\delta_{\ell}\right\rangle -\left\langle \delta_{i}\delta_{j}\right\rangle \left\langle \delta_{k}\delta_{\ell}\right\rangle \right].\label{eq:StartingPoint}
\end{equation}
To relate these quantities to the usual $n-$point functions, we must
rewrite the sum in terms of non-coincident points,
\begin{equation}
\sum_{i\neq j,k\neq\ell}\Theta_{ij}^{a}\Theta_{k\ell}^{b}n_{i}n_{j}n_{k}n_{\ell}\left\langle \delta_{i}\delta_{j}\delta_{k}\delta_{\ell}\right\rangle =\sum_{i\neq j\neq k\neq\ell}\Theta_{ij}^{a}\Theta_{k\ell}^{b}n_{i}n_{j}n_{k}n_{\ell}\left\langle \delta_{i}\delta_{j}\delta_{k}\delta_{\ell}\right\rangle +S_{\mathrm{single}}+S_{\mathrm{double}},
\end{equation}
with $S_{\mathrm{single}}$ containing single contractions,
\begin{eqnarray}
S_{\mathrm{single}} & = & \sum_{i\neq j\neq k}\Theta_{ij}^{a}\Theta_{ki}^{b}n_{i}^{2}n_{j}n_{k}\left\langle \delta_{i}^{2}\delta_{j}\delta_{k}\right\rangle \\
 & = & \sum_{i\neq j\neq k}\Theta_{ij}^{a}\Theta_{ki}^{b}n_{i}n_{j}n_{k}\left[\left\langle \delta_{j}\delta_{k}\right\rangle +\left\langle \delta_{i}\delta_{j}\delta_{k}\right\rangle \right],
\end{eqnarray}
and $S_{\mathrm{double}}$ containing double contractions,
\begin{eqnarray}
S_{\mathrm{double}} & = & \delta^{ab}\sum_{i\neq j}\Theta_{ij}^{a}n_{i}^{2}n_{j}^{2}\left\langle \delta_{i}^{2}\delta_{j}^{2}\right\rangle \\
 & = & \delta^{ab}\sum_{i\neq j}\Theta_{ij}^{a}n_{i}n_{j}\left[1+\left\langle \delta_{i}\delta_{j}\right\rangle \right].
\end{eqnarray}
We have used (\ref{eq:Contraction}) to simplify these expressions,
and $\delta^{ab}$ are Kronecker deltas. These contractions
can be visualized with the diagrams in fig. \ref{fig:Contractions}.

After considering the possible contractions, we arrive at the following
expression for $C^{ab}$, the model covariance matrix:
\begin{eqnarray}
C^{ab} & = & C_{4}^{ab}+C_{3}^{ab}+C_{2}^{ab}\,,\\
C_{4}^{ab} & \equiv & \left(RR^{a}RR^{b}\right)^{-1}\sum_{i\neq j\neq k\neq\ell}\Theta_{ij}^{a}\Theta_{k\ell}^{b}n_{i}n_{j}n_{k}n_{\ell}\left[\xi_{ijk\ell}+\xi_{ik}\xi_{j\ell}+\xi_{i\ell}\xi_{jk}\right],\\
C_{3}^{ab} & \equiv & 4\times\left(RR^{a}RR^{b}\right)^{-1}\sum_{i\neq j\neq k}\Theta_{ij}^{a}\Theta_{ki}^{b}n_{i}n_{j}n_{k}\left[\xi_{ijk}+\xi_{jk}\right],\\
C_{2}^{ab} & \equiv & 2\times\delta_{ab}\left(RR^{a}RR^{b}\right)^{-1}\sum_{i\neq j}\Theta_{ij}^{a}n_{i}n_{j}\left[\xi_{jk}+1\right],
\end{eqnarray}
where $\xi_{ijk}$ and $\xi_{ijk\ell}$ denote the connected 3- and
4-point functions, respectively. A continuum limit yields 
\begin{eqnarray}
C_{4}^{ab} & = & \left(RR^{a}RR^{b}\right)^{-1}\int d^{3}\vec{r}_{i}d^{3}\vec{r}_{j}d^{3}\vec{r}_{k}d^{3}\vec{r}_{\ell}\Theta^{a}\left(\vec{r}_{i}-\vec{r}_{j}\right)\Theta^{b}\left(\vec{r}_{k}-\vec{r}_{\ell}\right)n\left(\vec{r}_{i}\right)n\left(\vec{r}_{j}\right)n\left(\vec{r}_{k}\right)n\left(\vec{r}_{\ell}\right)\nonumber \\
 &  & \times\left[\xi_{4}\left(\vec{r}_{i},\vec{r}_{j},\vec{r}_{k},\vec{r}_{\ell}\right)+\xi\left(\vec{r}_{i}-\vec{r}_{k}\right)\xi\left(\vec{r}_{j}-\vec{r}_{\ell}\right)+\xi\left(\vec{r}_{i}-\vec{r}_{k}\right)\xi\left(\vec{r}_{j}-\vec{r}_{\ell}\right)\right],\\
C_{3}^{ab} & = & 4\times\left(RR^{a}RR^{b}\right)^{-1}\int d^{3}\vec{r}_{i}d^{3}\vec{r}_{j}d^{3}\vec{r}_{k}\Theta^{a}\left(\vec{r}_{i}-\vec{r}_{j}\right)\Theta^{b}\left(\vec{r}_{k}-\vec{r}_{i}\right)n\left(\vec{r}_{i}\right)n\left(\vec{r}_{j}\right)n\left(\vec{r}_{k}\right)\nonumber \\
 &  & \times\left[\xi_{3}\left(\vec{r}_{i},\vec{r}_{j},\vec{r}_{k}\right)+\xi\left(\vec{r}_{i}-\vec{r}_{k}\right)\right],\\
C_{2}^{ab} & = & 2\delta_{ab}\times\left(RR^{a}RR^{b}\right)^{-1}\int d^{3}\vec{r}_{i}d^{3}\vec{r}_{j}\Theta^{a}\left(\vec{r}_{i}-\vec{r}_{j}\right)n\left(\vec{r}_{i}\right)n\left(\vec{r}_{j}\right)\nonumber \\
 &  & \times\left[\xi\left(\vec{r}_{i}-\vec{r}_{j}\right)+1\right],\\
RR^{a} & = & \int d^{3}\vec{r}_{i}d^{3}\vec{r}_{j}\Theta^{a}\left(\vec{r}_{i}-\vec{r}_{j}\right)n\left(\vec{r}_{i}\right)n\left(\vec{r}_{j}\right).
\end{eqnarray}
with $\xi_{3}$ and $\xi_{4}$ now denoting the connected 3- and 4-point
functions and $n\left(\vec{r}\right)$ the expected number density
at $\vec{r}$. 

We can introduce a final complication, which is a non-uniform weighting
scheme (e.g. FKP weights \citep{Feldman:1993ky}). The basic observation
is that a factor of $w_{i}w_{j}$ is added to (\ref{eq:xi_estimator})
and (\ref{eq:rr_def}), leading to an additional factor
of $w_{i}w_{j}w_{k}w_{\ell}$ in (\ref{eq:StartingPoint}), but that
(\ref{eq:Contraction}) is unmodified. We therefore find
\begin{eqnarray}
C_{4}^{ab} & = & \left(RR^{a}RR^{b}\right)^{-1}\int d^{3}\vec{r}_{i}d^{3}\vec{r}_{j}d^{3}\vec{r}_{k}d^{3}\vec{r}_{\ell}\Theta^{a}\left(\vec{r}_{i}-\vec{r}_{j}\right)\Theta^{b}\left(\vec{r}_{k}-\vec{r}_{\ell}\right)\nonumber \\
 &  & \times n\left(\vec{r}_{i}\right)n\left(\vec{r}_{j}\right)n\left(\vec{r}_{k}\right)n\left(\vec{r}_{\ell}\right)w\left(\vec{r}_{i}\right)w\left(\vec{r}_{j}\right)w\left(\vec{r}_{k}\right)w\left(\vec{r}_{\ell}\right)\\
 &  & \times\left[\xi_{4}\left(\vec{r}_{i},\vec{r}_{j},\vec{r}_{k},\vec{r}_{\ell}\right)+\xi\left(\vec{r}_{i}-\vec{r}_{k}\right)\xi\left(\vec{r}_{j}-\vec{r}_{\ell}\right)+\xi\left(\vec{r}_{i}-\vec{r}_{\ell}\right)\xi\left(\vec{r}_{j}-\vec{r}_{k}\right)\right],\label{eq:C4}\\
C_{3}^{ab} & = & 4\times\left(RR^{a}RR^{b}\right)^{-1}\int d^{3}\vec{r}_{i}d^{3}\vec{r}_{j}d^{3}\vec{r}_{k}\Theta^{a}\left(\vec{r}_{i}-\vec{r}_{j}\right)\Theta^{b}\left(\vec{r}_{k}-\vec{r}_{i}\right)n\left(\vec{r}_{i}\right)n\left(\vec{r}_{j}\right)n\left(\vec{r}_{k}\right)\nonumber \\
 &  & \times w^{2}\left(\vec{r}_{i}\right)w\left(\vec{r}_{j}\right)w\left(\vec{r}_{k}\right)\left[\xi_{3}\left(\vec{r}_{i},\vec{r}_{j},\vec{r}_{k}\right)+\xi\left(\vec{r}_{i}-\vec{r}_{k}\right)\right],\\
C_{2}^{ab} & = & 2\delta_{ab}\times\left(RR^{a}RR^{b}\right)^{-1}\int d^{3}\vec{r}_{i}d^{3}\vec{r}_{j}\Theta^{a}\left(\vec{r}_{i}-\vec{r}_{j}\right)n\left(\vec{r}_{i}\right)n\left(\vec{r}_{j}\right)\nonumber \\
 &  & \times w^{2}\left(\vec{r_{i}}\right)w^{2}\left(\vec{r_{j}}\right)\left[\xi\left(\vec{r}_{i}-\vec{r}_{j}\right)+1\right],\label{eq:C2}\\
RR^{a} & = & \int d^{3}\vec{r}_{i}d^{3}\vec{r}_{j}\Theta^{a}\left(\vec{r}_{i}-\vec{r}_{j}\right)n\left(\vec{r}_{i}\right)n\left(\vec{r}_{j}\right)w\left(\vec{r_{i}}\right)w\left(\vec{r_{j}}\right).\label{eq:RR}
\end{eqnarray}
$C_{2}$, $C_{3}$, and $C_{4}$ all scale as $1/\mathrm{Vol}$, and
are independent of overall rescalings of the weights. They also exhibit
a simple scaling with the number density
\begin{equation}
C_{m}^{ab}\propto n^{m-4}\,.\label{eq:n-scaling}
\end{equation}
In sec. \ref{sec:Integrate} we will numerically integrate the \emph{Gaussian}
terms in (\ref{eq:C4})-(\ref{eq:C2}). Rather than attempt to integrate
the \emph{non-Gaussian }terms involving $\xi_{3}$ and $\xi_{4}$,
in sec. \ref{sec:NGModel} we will develop a simple method that uses
increased shot noise to approximate the effects of non-Gaussianity.

\section{Integrating the Gaussian Model}
\label{sec:Integrate}

In order to test the value of equations (\ref{eq:C4})-(\ref{eq:RR})
for generating covariance matrices, we will attempt to reproduce the
covariance matrix of a set of mock catalogs. Specifically, we will
use 1,000 quick particle mesh (QPM) mocks \citep{White:2013psd} that
mimic the CMASS sample of the Baryon Oscillation Spectroscopic
Survey (BOSS). Redshift space distortions (RSD) are included, but
no reconstruction algorithms {[}REFS{]} have been applied to these
mocks.

We use a correlation function estimated in 35 radial bins of width
$\delta r=4\, h^{-1}\mathrm{Mpc}$, covering $r=40-180\, h^{-1}\mathrm{Mpc}$,
and 10 angular bins of width $\delta\mu=0.1$, covering $\mu=0-1$.
This is more bins, by at least a factor of 5, than would be used in
a typical  multipole analysis of BAO in the galaxy correlation function.
A more aggressive binning was chosen in part because it makes the
covariance matrix problem more challenging and makes the noise in
the sample covariance matrix more readily apparent.

The numerical integration is performed with a new software package%
\footnote{An introduction to the package, as well as the python code, will be
available at a later date.%
}, ``Rascal: A Rapid Sampler for Large Covariance Matrices''. It
allows flexible specification of the cosmology, redshift distribution,
and model correlation function, and can accept angular completeness
masks in the MANGLE format \citep{Swanson:2007aj}. The sampling algorithm
is described below, in sec. \ref{sub:Sampling}.

\subsection{Survey Geometry}

The QPM mocks have a flat $\Lambda\mathrm{CDM}$ cosmology with $\Omega_{m}=0.29$.
The geometry of the mock catalogs can be decomposed into radial and
angular components. In the radial direction, the redshift-dependent
number density $\overline{n}\left(z\right)$ and weights $w\left(z\right)$
are calibrated to the QPM mocks. We set $\overline{n}\left(z\right)=0$
outside of the redshift bounds of the CMASS sample, i.e. for $z<0.43$
and $z>0.7$. In the angular directions, the survey mask specifies
the completeness of the survey, $c\left(\Omega\right)$. The number
density at a point in the survey is thus given by 
\begin{equation}
n\left(z,\Omega\right)=\overline{n}\left(z\right)c\left(\Omega\right).
\end{equation}

\begin{figure}
\includegraphics{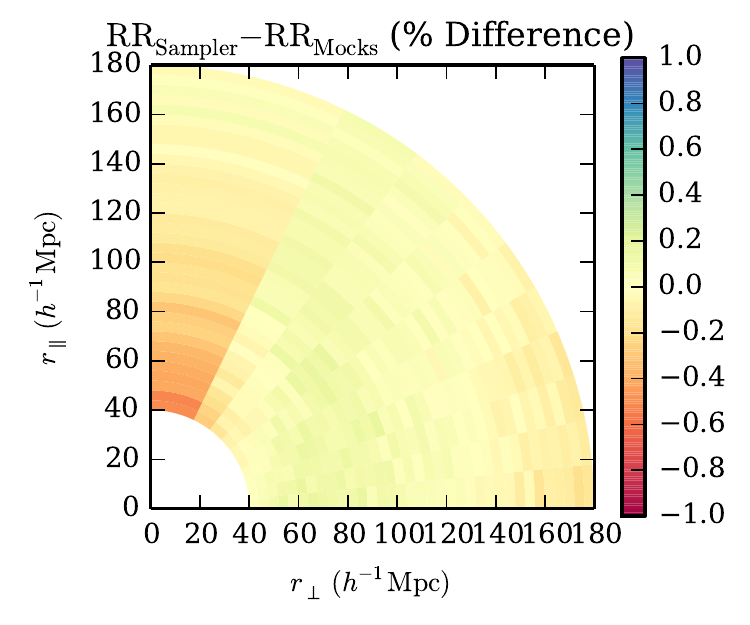}\hfill{}\includegraphics{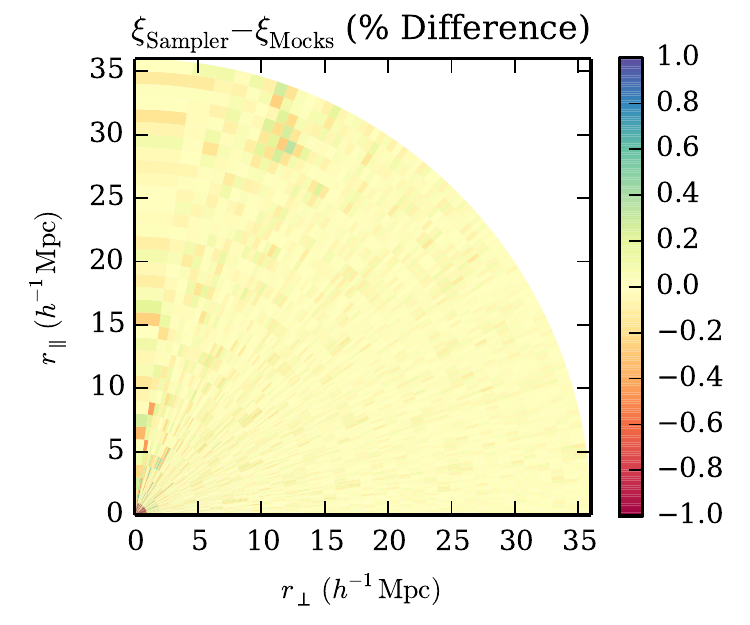}

\caption{\label{fig:SurveyComparison}Using RR pair counts as a test of the
survey geometry, we see that the representation of the survey in our
sampler is in good agreement with the survey geometry in the mocks.
We test the correlation function in our sampler by integrating it
over the same bins used in the mocks, and find the two to be in excellent
agreement.}
\end{figure}

In order to compare our implementation of the survey geometry with
that of the mock catalogs, we compare the values of $RR^{a}$ that
we compute with the values taken from the mock random catalogs. The
results are shown in fig. \ref{fig:SurveyComparison}. We find a small
offset between our integration and the mocks which varies as a function
of $r_{\perp}$. This arises because the mocks are generated using
a ``veto mask'' which replicates holes in the survey smaller than
the telescope field-of-view (e.g. areas obscured by bright stars).
Querying this mask is quite time-consuming, and we have not used in
when generating the samples in this paper.

\subsection{Correlation Function}

In order to integrate equations (\ref{eq:C4})-(\ref{eq:C2}) we must
specify a model correlation function (not to be confused with the
estimated correlation function, whose covariance matrix we are computing).
Rather than use the linear theory correlation function, or one derived
from perturbation theory, we use the non-linear correlation function
estimated from the 1,000 QPM mocks. This ensures that the correlation
function we use for integration agrees at relatively short scales
($r\gtrsim1\, h^{-1}\mathrm{Mpc}$) with the mocks, faithfully reproducing
non-linear features such as the ``finger-of-God''.

In the mocks, the correlation function is evaluated with radial bins
that are $1\, h^{-1}\mathrm{Mpc}$ wide and 120 angular bins covering
$\mu=0-1$. A potential problem is that we then have estimates of
the correlation function \emph{averaged over }a bin, while the interpolation
scheme used in our code assumes that the correlation function is specified
\emph{at the center} of the bin. In order to address this we adopted
an iterative scheme. We first treated the bin-averaged values $\xi_{\mathrm{mock}}^{m}$
from the mocks as bin-center values, then performed the analogous
bin-averaged integrals using our code,
\begin{equation}
\xi_{\mathrm{out}}^{m}=\left(RR^{a}\right)^{-1}\int d\vec{r}_{i}d\vec{r}_{j}\Theta^{m}\left(\vec{r}_{i}-\vec{r}_{j}\right)n\left(\vec{r}_{i}\right)w\left(\vec{r}_{i}\right)n\left(\vec{r}_{j}\right)w\left(\vec{r}_{j}\right)\xi_{in}\left(\vec{r}_{i}-\vec{r}_{j}\right).
\end{equation}
We then rescaled the bin-centered input values of $\xi$ according
to $\xi_{\mathrm{out}}^{m}/\xi_{\mathrm{mock}}^{m}$ and repeated
the procedure. After three iterations, we arrived at sub-percent agreement
between our implementation of the correlation function, as measured
by its bin-averages, and the correlation function in the mocks. This
is demonstrated in fig. \ref{fig:SurveyComparison}.

\subsection{\label{sub:Sampling}Importance Sampling}

The integral (\ref{eq:C4}) is 12-dimensional. It also has relatively
limited support, so that the simplest Monte Carlo approaches converge
quite slowly. We achieve acceptable convergence rates by using importance
sampling. We begin by choosing a bin $\Theta^{a}\left(\vec{r}_{i}-\vec{r}_{j}\right)$,
then generate uniform draws of the separation $\vec{r}_{i}-\vec{r}_{j}$
within that bin. We then draw $\left|\vec{r}_{j}-\vec{r}_{k}\right|$
and $\left|\vec{r}_{i}-\vec{r}_{\ell}\right|$ from a pdf proportional
to $r^{2}\left|\xi_{0}\left(r\right)\right|$, where $\xi_{0}\left(r\right)$
is the spherically-averaged correlation function. The angular components
of $\vec{r}_{j}-\vec{r}_{k}$ and $\vec{r}_{i}-\vec{r}_{\ell}$ are
again determined by uniform draws. Finally, we sort the sets
of four points according to which bin $\Theta^{b}\left(\vec{r}_{k}-\vec{r}_{\ell}\right)$
they fall in. We therefore are able to estimate one column of $C_{4}^{ab}$
from a set of draws. Since the integration proceeds column-by-column,
this approach is trivially parallelizable. Moreover, the same draws
can be used to determine $C_{3}^{ab}$ and $C_{2}^{aa}$. 

\begin{figure}
\includegraphics{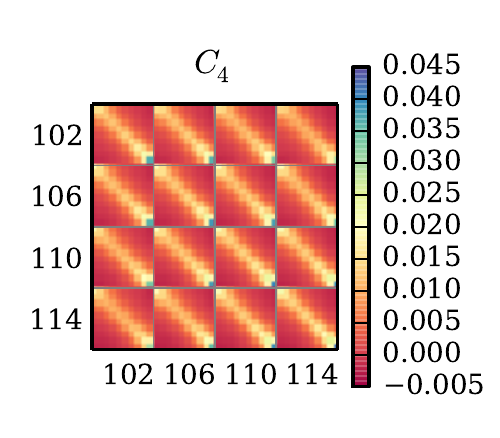}\hfill{}\includegraphics{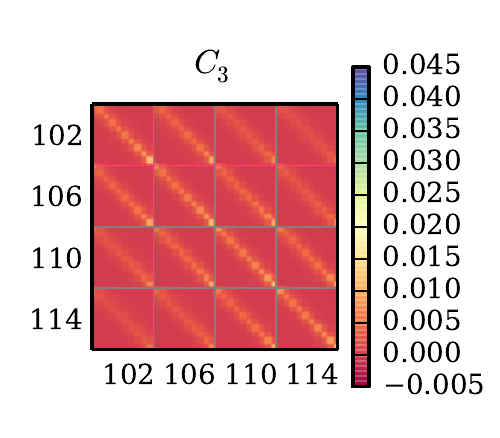}\hfill{}\includegraphics{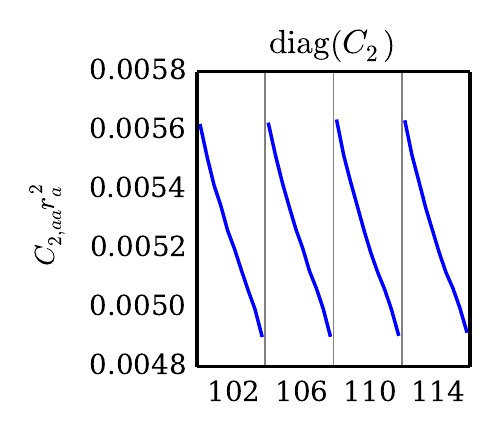}

\caption{\label{fig:Templates}The three covariance matrix templates defined
in (\ref{eq:C4})-(\ref{eq:C2}), as evaluated using our sampler.
The Gaussian covariance matrix is the sum of $C_{4}$, $C_{3}$, and
the diagonal matrix $C_{2}$. We show only a small portion of the
covariance matrix, but this is sufficient to illustrate the high degree
of convergence achieved in $\sim1,000$ CPU hours with our sampler.}

\end{figure}

In fig. \ref{fig:Templates} we show portions of the $C_{4}$, $C_{3}$,
and $C_{2}$ that we computed using this method and the inputs described
above. These runs were completed on a desktop computer, requiring
$\approx1,000$ CPU hours to complete. Although the noise in our samples
is not Wishart distributed, we can characterize our results in terms
of an effective number of mocks by choosing a range of bins where
the precision matrix should be very close to zero, then measuring
the noise in those bins. Inverse Wishart statistics would imply 
\begin{equation}
\mathrm{var}\left(\Psi^{ab}\right)\approx\frac{\Psi^{aa}\Psi^{bb}}{N_{\mathrm{mocks}}-N_{\mathrm{bins}}}
\end{equation}
for bins where $\Psi^{ab}\approx0$. We therefore define the effective
number of mocks as 
\begin{equation}
N_{\mathrm{eff}}\equiv N_{\mathrm{bins}}+\left[\mathrm{var}\left(\frac{\Psi^{ab}}{\sqrt{\Psi^{aa}\Psi^{bb}}}\right)\right]^{-1},
\end{equation}
with the variance taken across bins with $\Psi^{ab}\approx0$ from
a single realization. Using bins with $r^{a}\ge142\, h^{-1}\mathrm{Mpc}$
and $r^{b}\le82\, h^{-1}\mathrm{Mpc},$ we find $N_{\mathrm{eff}}\sim35,000$
for the $C_{4}$, $C_{3}$, and $C_{2}$ generated for this paper.
The convergence of the sampler is easily characterized in terms of
$N_{\mathrm{eff}}$ , with a roughly linear relationship between CPU
time used running the sampler and $N_{\mathrm{eff}}$. 

\subsection{\label{sub:Comparison-with-Mocks}Comparison with Mocks}

\begin{figure}
\includegraphics{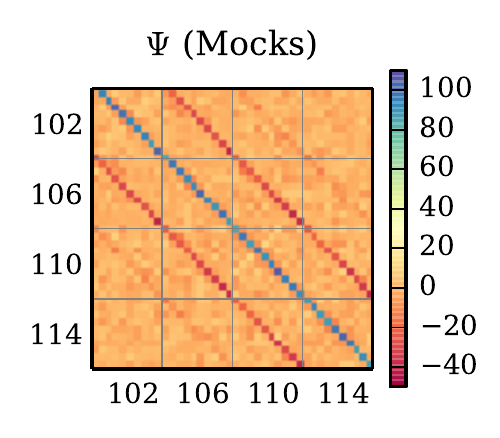}\hfill{}\includegraphics{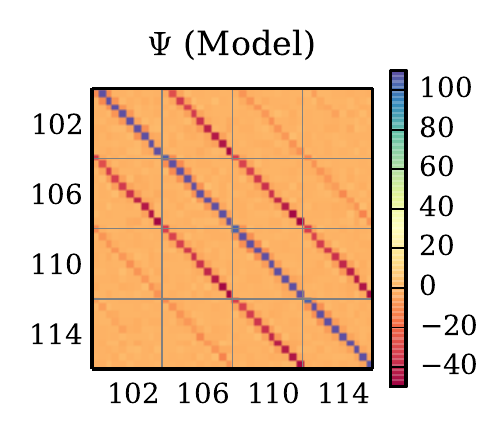}\hfill{}\includegraphics{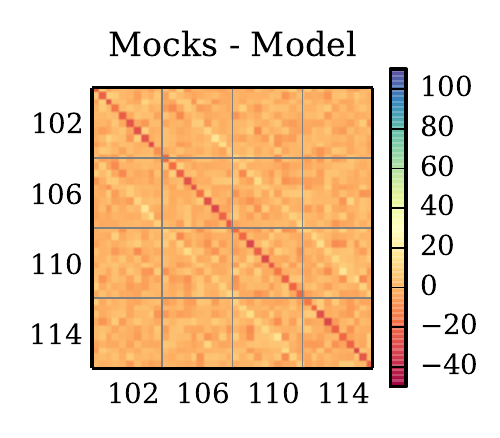}

\caption{\label{fig:UnfitPrecBoxes}A small portion of the precision (inverse
covariance) matrix, as determined from mock catalogs and from our
Gaussian model (without fitting). The correlation function is evaluated
in bins with $\delta r=4\, h^{-1}\mathrm{Mpc}$ and $\delta\mu=0.1$.
In these plots bins are ordered first in $r$ (covering $r=102\, h^{-1}\mathrm{Mpc}$
to $r=114\, h^{-1}\mathrm{Mpc}$), then in $\mu$ ($\mu=0$ to $\mu=1$).
The reduction in noise in the model precision matrix, relative to
the mock precision matrix, is readily apparent. The residuals in the
rightmost plot are surprisingly small, given that we are using the
purely Gaussian model, but also indicate the need for non-Gaussian
corrections to the model (see fig. \ref{fig:PrecBoxes}).}
\end{figure}

We have constructed a simple Gaussian model for the covariance matrix,
taking as inputs the survey geometry and non-linear correlation function.
Given the modesty of these inputs and the simplicity of the Gaussian
model, the close agreement between our model covariance matrix and
the covariance matrix determined using 1,000 QPM mock catalogs with
the same geometry and correlation function is remarkable. In figure
\ref{fig:UnfitPrecBoxes} we plot a portion of the precision matrix
as determined from the mocks, as computed in the Gaussian model, and
the difference between the two. We observe discrepancies between the
two at the $\sim10\%$ level, most noticeable for diagonal entries
in the precision matrix and in entries for bins with the same $\mu$
that are adjacent in $r$, and we now investigate the consequences
of these discrepancies.

There are two reasons for using the precision matrix rather than the
covariance matrix in this comparison. First, the structure of the
precision matrix is considerably simpler than the structure of the
covariance matrix, as can be seen by comparing fig. \ref{fig:Templates}
and fig. \ref{fig:UnfitPrecBoxes}. Second, the precision matrix,
rather than the covariance matrix, is what is used to compute $\chi^{2}$
statistics, and thus is the quantity that is most relevant in analysis.
Finally, we point out that large eigenvalues of the precision matrix
correspond to small eigenvalues of the covariance matrix, and vice
versa, so agreement between precision matrices cannot easily be established
by comparing the corresponding covariance matrices directly.

\begin{figure}
\includegraphics{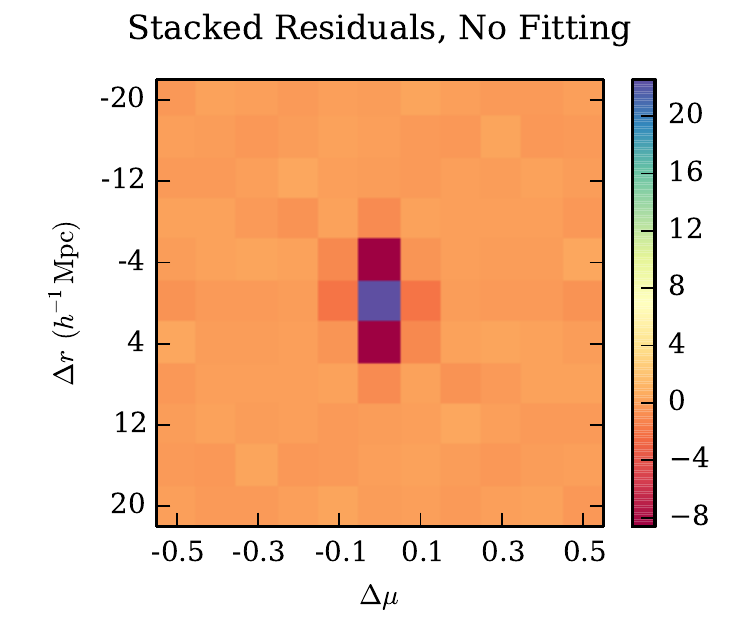}\hfill{}\includegraphics{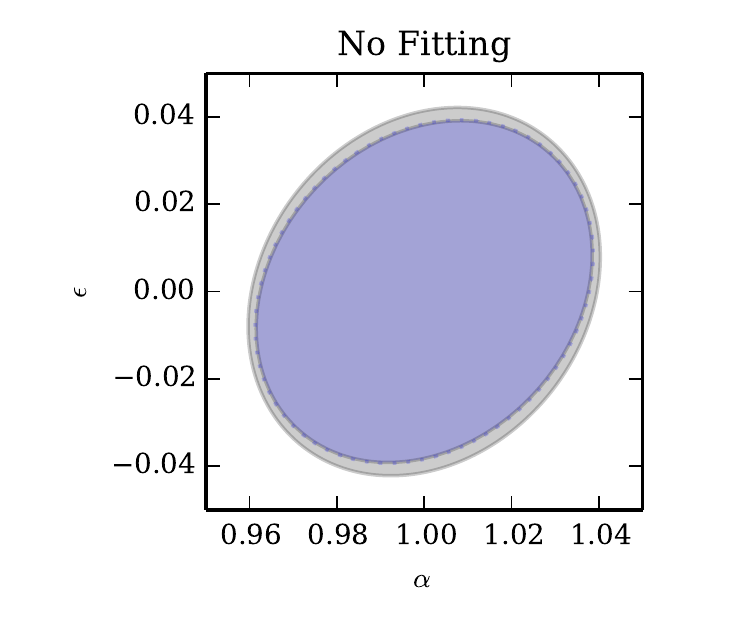}

\caption{\label{fig:GaussianFitting}Comparison between the mock precision
matrix and our (unfit) Gaussian model precision. Stacking the residuals
reveals discrepancies on the diagonal of the precision matrix, and
also for bins that share a $\mu$ range and are adjacent in $r$.
Error ellipses for the BAO parameters $\alpha$ and $\epsilon$ show
small discrepancies between the mocks (gray), model (blue), and noisy
realizations of the model (dashed). The purely Gaussian model performs
surprisingly well for BAO error estimation, but the discrepancies
can be reduced with a non-Gaussian extension of the model (see fig.
\ref{fig:FitEllipses}).}
\end{figure}

We have chosen two approaches to a more detailed comparison of the
precision matrices. In the first approach we begin by computing
weighted residuals between the two precision matrices, 
\begin{equation}
\mathrm{resid}^{ab}=\frac{\Psi_{\mathrm{model}}^{ab}-\Psi_{\mathrm{mock}}^{ab}}{r_{a}r_{b}}\,,
\end{equation}
with the $r-$weighting chosen to remove the naive $r-$dependence
of the precision matrix. We then compute for each bin 
\begin{eqnarray}
\Delta r^{ab} & = & r^{a}-r^{b}\,,\\
\Delta\mu^{ab} & = & \mu^{a}-\mu^{b},
\end{eqnarray}
and average together the residuals for all bins with the same values
of $\Delta r$ and $\Delta\mu$. The result is shown in fig. \ref{fig:GaussianFitting},
where it is clear that we observe significant discrepancies when $\Delta r=\Delta\mu=0$
and when $\Delta r=4\, h^{-1}\mathrm{Mpc}$, $\Delta\mu=0$. We also
observe small residuals for bins with $\Delta r=0\, h^{-1}\mathrm{Mpc}$,
$\Delta\mu=0.1$, which were not readily apparent in figure \ref{fig:UnfitPrecBoxes}.

While the first method of comparison provides a direct check on the
precision matrix, at least for modest $\Delta r$ and $\Delta\mu$,
the second method uses a Fisher matrix to focus on modes of interest
in BAO analyses. We assume a simple model for the correlation function,
\begin{equation}
\xi_{\mathrm{Fisher}}\left(r,\mu\right)=\xi_{0}\left(b,\beta,\alpha,\epsilon;r\right)P_{0}\left(\mu\right)+\xi_{2}\left(b,\beta,\alpha,\epsilon;r\right)P_{2}\left(\mu\right)+\sum_{i=0}^{2}\left(a_{0,i}r^{i-2}P_{0}\left(\mu\right)+a_{2,i}r^{i-2}P_{2}\left(\mu\right)\right).
\end{equation}
This correlation function depends on ten parameters: the correlation
function bias $b$, redshift-space distortion parameter $\beta$,
isotropic BAO rescaling factor $\alpha$, anisotropic BAO rescaling
factor $\epsilon$, and the six systematic parameters $a_{0,i},a_{2,i}$.
$P_{0}$ and $P_{2}$ are Legendre polynomials. We then use the usual
Fisher matrix formula to relate the precision matrix for these parameters
to the precision matrix for the correlation function:
\begin{equation}
\Psi_{\mathrm{Parameters}}^{ij}=\frac{\partial\xi_{\mathrm{Fisher}}^{T}}{\partial p_{i}}\Psi_{\mathrm{\xi}}\frac{\partial\xi_{\mathrm{Fisher}}}{\partial p_{j}}\,,
\end{equation}
with $p_{i}$ running over the ten parameters. We invert $\Psi_{\mathrm{Parameters}}$
to find $C_{\mathrm{Parameters}}$, then marginalize over $b$, $\beta$,
and the systematic parameters to find a covariance matrix for $\alpha$
and $\epsilon$ alone. The resulting error ellipses are shown in fig.
\ref{fig:GaussianFitting}, and the parameters that define those ellipses
are show in table \ref{tab:Ellipses}. Again, discrepancies are readily
apparent, though we defer a quantitative discussion of those discrepancies
for sec. \ref{sub:Results}.

It is telling that the discrepancies we observe between the mock and
model precision matrices appear primarily on the diagonal of the precision
matrix, and for bins that are adjacent to one another. Referring back
to the integral in (\ref{eq:C4}), we see that e.g. $\left|\vec{r}_{i}-\vec{r}_{k}\right|$
and $\left|\vec{r}_{j}-\vec{r}_{\ell}\right|$ can only reach very
small values for overlapping or adjacent bins. We therefore interpret
these discrepancies as a sign that our model is missing a contribution
at short scales; since we know the 2-point correlation function at
short scales very well, it seems most likely that the observed discrepancies
are a consequence of dropping the non-Gaussian terms involving $\xi_{3}$
and $\xi_{4}$. In the following we will attempt to model the non-Gaussian
contributions without building explicit models for $\xi_{3}$ and
$\xi_{4}$.

\section{A Simple Model for Non-Gaussianity}
\label{sec:NGModel}

We now attempt to improve the the results of the previous section
by introducing a one-parameter model to accommodate the effects of
non-Gaussianity. Roughly speaking, we expect the connected 3- and
4-point functions to be large at small separations, then fall off
rapidly at larger separations. A simple way of making the galaxy field
``more correlated'' at short scales is to increase the amount of
shot noise. Recall from (\ref{eq:n-scaling}) that $C_{m}\propto n^{m-4}$,
so that we can implement this rescaling using the integrals performed
in the previous section:
\begin{equation}
C_{\mathrm{NG}}\left(a\right)=C_{4}+aC_{3}+a^{2}C_{2}\,.\label{eq:C_NG}
\end{equation}
As we will show, this simple rescaling yields a precision matrix that
is in excellent agreement with the mock precision matrix. For applications
of a similar model to the covariance of the power spectrum, see \citep{Carron:2014hja}.

In the following we will develop two approaches for fitting $C_{\mathrm{NG}}$
to the mocks. While the two approaches have distinct advantages and
disadvantages, they will lead to substantially similar error ellipses
for the BAO parameters $\alpha$ and $\epsilon$.

\subsection{$\mathcal{L}_{1}$ Likelihood}

If we consider the mock correlation functions as noisy draws from
a distribution described by $\Psi_{\mathrm{NG}}\left(a\right)=C_{\mathrm{NG}}^{-1}\left(a\right)$,
the likelihood for $a$ is 
\begin{eqnarray}
\mathcal{L}_{1}\left(a\right) & = & \prod_{\mathrm{\alpha\in mocks}}\sqrt{\frac{\det\Psi_{\mathrm{NG}}\left(a\right)}{\left(2\pi\right)^{N_{\mathrm{bins}}}}}\exp\left[-\frac{1}{2}\xi_{\alpha}^{T}\Psi_{\mathrm{NG}}\left(a\right)\xi_{\alpha}\right].
\end{eqnarray}
With a bit of rearranging, this becomes 
\begin{equation}
-\log\mathcal{L}_{1}\left(a\right)=\frac{N_{\mathrm{mocks}}}{2}\left[\mathrm{tr}\left(\Psi_{\mathrm{NG}}\left(a\right)C_{\mathrm{mocks}}\right)-\log\det\Psi_{\mathrm{NG}}\left(a\right)\right]+\dots\,,\label{eq:L1}
\end{equation}
where the omitted terms are independent of $a$. Given our expression
(\ref{eq:C_NG}) for the model covariance matrix and the mock covariance
matrix, it is straightforward to find the maximum likelihood value
for $a$. This fitting dramatically reduces the residuals between
the model and mock precision matrices, as shown in fig. \ref{fig:FitResiduals},
and brings the mock and model $\alpha-\epsilon$ error ellipses into
better agreement, as show in fig. \ref{fig:FitEllipses}. 

In order to better understand how this fitting works, we expand around
a solution $\Psi_{\mathrm{NG}}\to\Psi_{\mathrm{NG}}+\delta\Psi_{\mathrm{NG}}$,
finding 
\begin{eqnarray}
\delta\left[-\log\mathcal{L}_{1}\right] & = & \frac{N_{\mathrm{mocks}}}{2}\left[\mathrm{tr}\left(\delta\Psi_{\mathrm{NG}}C_{\mathrm{mocks}}\right)-\log\det\left(1+C_{\mathrm{NG}}\delta\Psi_{\mathrm{NG}}\right)\right]\\
 & \approx & \frac{N_{\mathrm{mocks}}}{2}\mathrm{tr}\left[\delta\Psi_{\mathrm{NG}}\left(C_{\mathrm{mocks}}-C_{\mathrm{NG}}\right)\right],
\end{eqnarray}
so that the likelihood $\mathcal{L}_{1}$ is maximized when $C_{\mathrm{mocks}}=C_{\mathrm{NG}}$.

The principal benefit of the likelihood $\mathcal{L}_{1}$ is that
it does not require that we invert $C_{\mathrm{mocks}}$. Indeed,
this likelihood provides effective constraints on $a$ (or other parameters,
should we choose to fit a more elaborate model) even when $C_{\mathrm{mocks}}$
is degenerate, i.e. when $N_{\mathrm{mocks}}<N_{\mathrm{bins}}$.
Although we will continue to use the full 1,000 QPM mocks in our analysis,
a much smaller number would have led to a very similar $\Psi_{\mathrm{NG}}$. We anticipate that because this approach alleviates the need to have
very large numbers of mock catalogs, it could be useful in a wide
variety of cosmological analyses.

\subsection{$\mathcal{L}_{2}$ Likelihood}

The $\mathcal{L}_{1}$ is not the only possible way to fit one covariance
matrix to another. If we consider the goal to be comparison of two
matrices, a plethora of methods are available. Noting that our goal
is not to compare to arbitrary matrices, but instead to compare the
two \emph{distributions} described by those matrices, significantly
narrows the field. One prominent candidate is the Kullback-Leibler
(KL) divergence \citep{Kullback:1951aa}, which for two multivariate
normal distributions (with the same means) specified by $\Psi_{a}$
and $C_{b}$ simplifies to 
\begin{equation}
KL\left(\Psi_{a},C_{b}\right)=\mathrm{tr}\left(\Psi_{a}C_{b}\right)-\log\det\Psi_{a}-\log\det C_{b}-N_{\mathrm{bins}}\,.
\end{equation}
Note that if we take $\Psi_{a}=\Psi_{\mathrm{NG}}\left(a\right)$
and $C_{b}=C_{\mathrm{mocks}}$, the result is proportional to (\ref{eq:L1}),
so that we have arrived back at the $\mathcal{L}_{1}$ likelihood.

\begin{figure}
\includegraphics{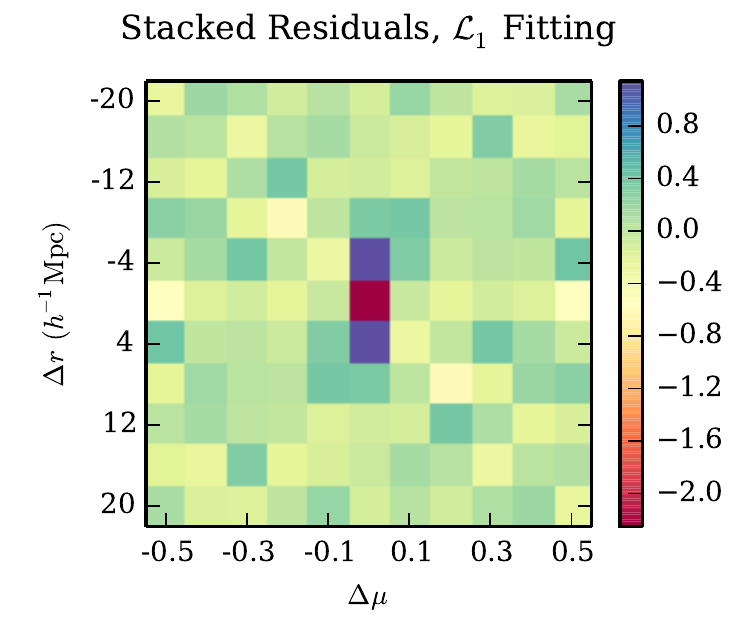}\hfill{}\includegraphics{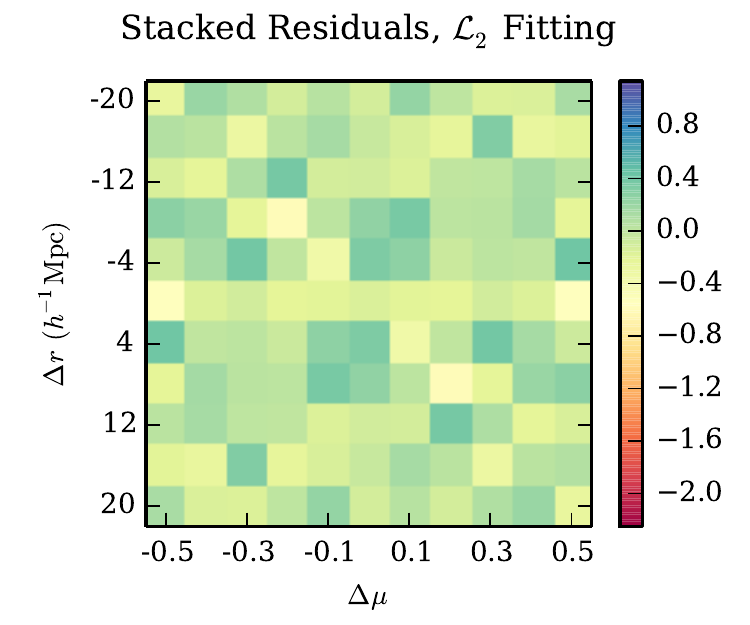}

\caption{\label{fig:FitResiduals}Stacked residuals between our non-Gaussian
(fit) models and the mocks. Fitting the non-Gaussian model using the
$\mathcal{L}_{1}$ likelihood significantly reduces the stacked residuals
relative to the (unfit) Gaussian model (see fig. \ref{fig:GaussianFitting}),
while fitting with the $\mathcal{L}_{2}$ likelihood effectively eliminates
the residuals. The absence of residuals indicates that the model precision
matrix could replace the mock precision matrix in a variety of applications.}
\end{figure}

The KL divergence is notoriously non-symmetric, so that $KL\left(\Psi_{a},C_{b}\right)\neq KL\left(\Psi_{b},C_{a}\right)$.
We can therefore formulate a second likelihood,
\begin{equation}
-\log\mathcal{L}_{2}\left(a\right)=KL\left(\Psi_{\mathrm{mocks}},C_{\mathrm{NG}}\left(a\right)\right),
\end{equation}
which will lead to a slightly different maximum likelihood value for
$a$. Varying this likelihood as above, we find 
\begin{equation}
\delta\left[-\log\mathcal{L}_{2}\right]=\mathrm{tr}\left[\delta C_{\mathrm{NG}}\left(\Psi_{\mathrm{mocks}}-\Psi_{\mathrm{NG}}\right)\right],
\end{equation}
so that where $\mathcal{L}_{1}$ required agreement between $C_{\mathrm{mocks}}$
and $C_{\mathrm{NG}}$, $\mathcal{L}_{2}$ requires agreement between
$\Psi_{\mathrm{mocks}}$ and $\Psi_{\mathrm{NG}}$. While this should
not be an issue when the space of distributions spanned by $C_{\mathrm{NG}}\left(a\right)$
includes the distribution that gives rise to $C_{\mathrm{mocks}}$,
when this is \emph{not} the case we cannot expect $\mathcal{L}_{1}$
and $\mathcal{L}_{2}$ to yield the same maximum likelihood values.
Indeed, by examining fig. \ref{fig:FitResiduals}, which shows the
residuals between the mocks and the model precision matrixes, we can
see that the $\mathcal{L}_{1}$ leaves significant residuals while
$\mathcal{L}_{2}$ does not.

\begin{figure}
\includegraphics{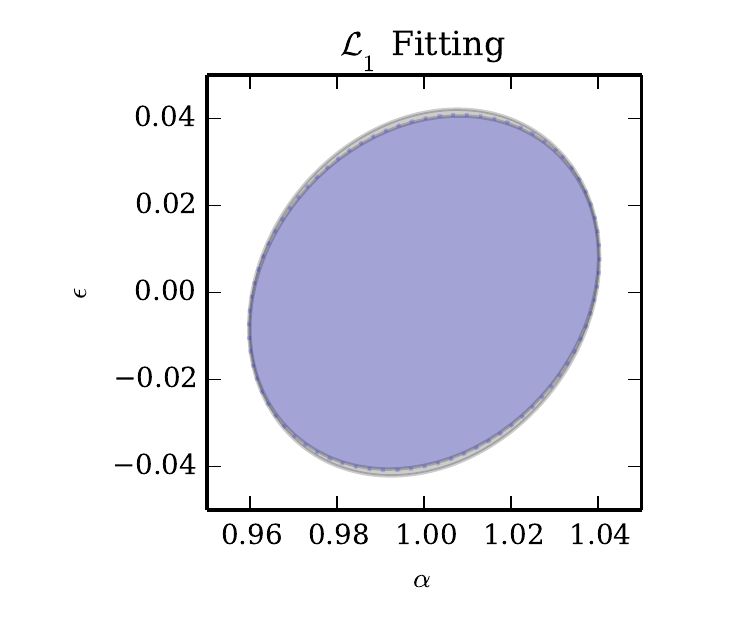}\hfill{}\includegraphics{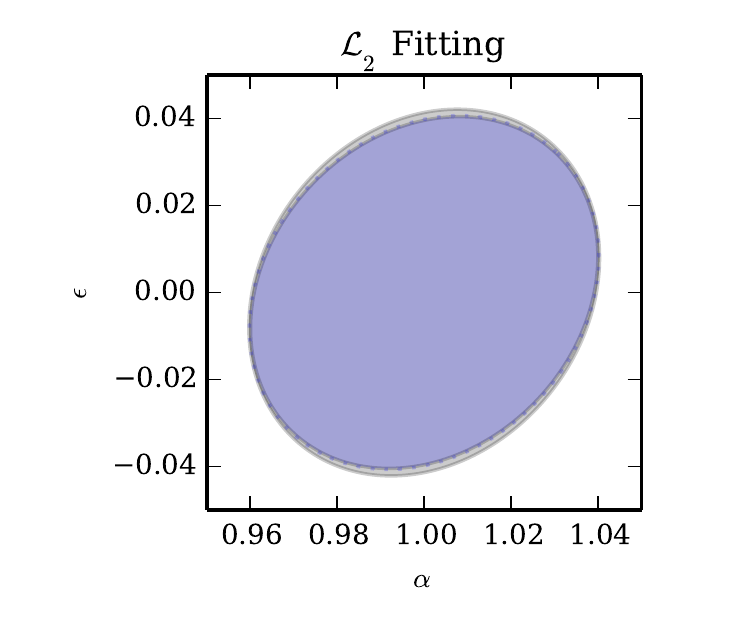}

\caption{\label{fig:FitEllipses}Error ellipses for the BAO parameters $\alpha$
and $\epsilon$ from the mocks (gray), the non-Gaussian (fit) models
(blue), and noisy realizations of those models (dashed). The discrepancies
between the mocks and the Gaussian (fit) model are reduced, though
a $3\%$ discrepancy persists in $\sigma_{\epsilon}$. The similarity
between the two error ellipses indicates that choice of likelihood
($\mathcal{L}_{1}$ or $\mathcal{L}_{2}$) for fitting the non-Gaussian
model has very little impact on BAO error estimation. }
\end{figure}

The principal benefit of the $\mathcal{L}_{2}$ likelihood is that
it produces better agreement between the mock and model precision
matrices than the $\mathcal{L}_{1}$ likelihood, as shown in fig.
\ref{fig:FitResiduals}. Note that most (but not all) applications
will require the precision matrix, rather than the covariance matrix.
The principal drawback is that it requires that we invert $C_{\mathrm{mocks}}$,
and so requires that $N_{\mathrm{mocks}}>N_{\mathrm{bins}}$. In our
case this requirement is satisfied, and so we prefer the $\mathcal{L}_{2}$
likelihood. This preference is however slight, as the resulting $\alpha-\epsilon$
error ellipses, shown in fig. \ref{fig:FitEllipses}, are quite similar.

\begin{figure}
\includegraphics{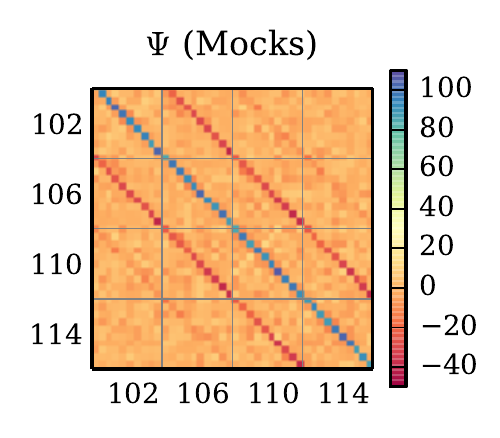}\hfill{}\includegraphics{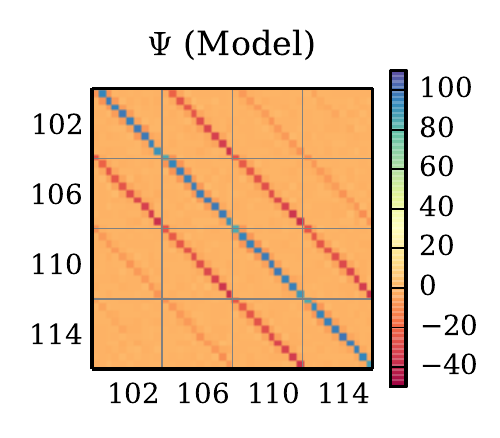}\hfill{}\includegraphics{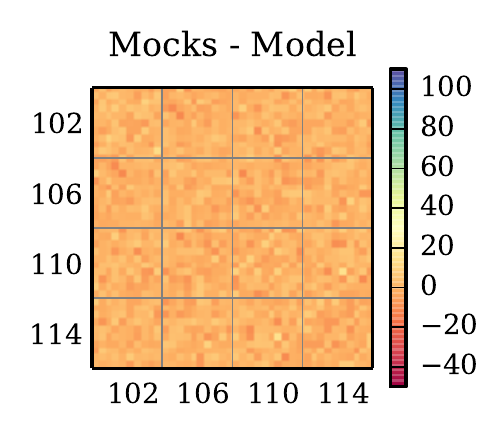}

\caption{\label{fig:PrecBoxes}A small portion of the precision matrix, as
determined from mock catalogs and from our non-Gaussian model (fit
using the $\mathcal{L}_{2}$ likelihood). The correlation function
is evaluated in bins with $\delta r=4\, h^{-1}\mathrm{Mpc}$ and $\delta\mu=0.1$.
In these plots bins are ordered first in $r$ (covering $r=102\, h^{-1}\mathrm{Mpc}$
to $r=114\, h^{-1}\mathrm{Mpc}$), then in $\mu$ ($\mu=0$ to $\mu=1$).
The difference in the final plot shows that fitting has removed the
residuals found in fig. \ref{fig:UnfitPrecBoxes}, so that the non-Gaussian
model precision matrix is now in excellent agreement with the mock
precision matrix.}
\end{figure}

\begin{figure}
\includegraphics{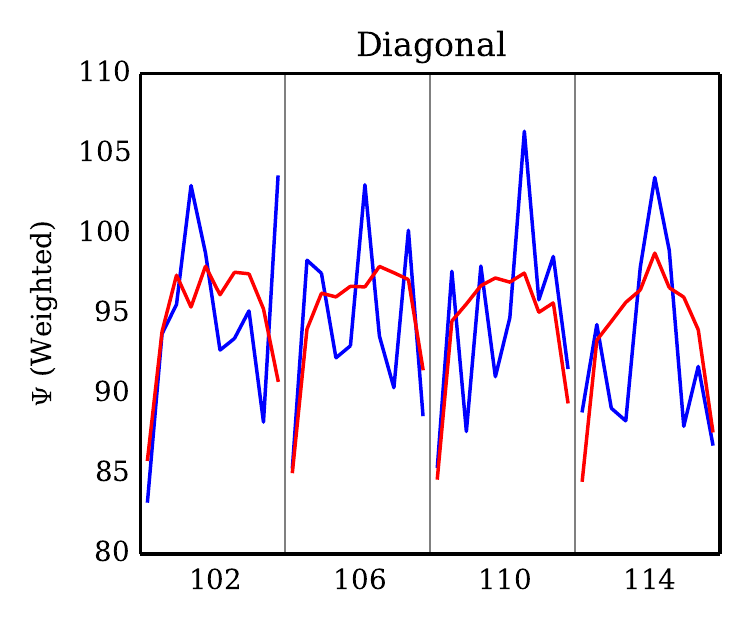}\hfill{}\includegraphics{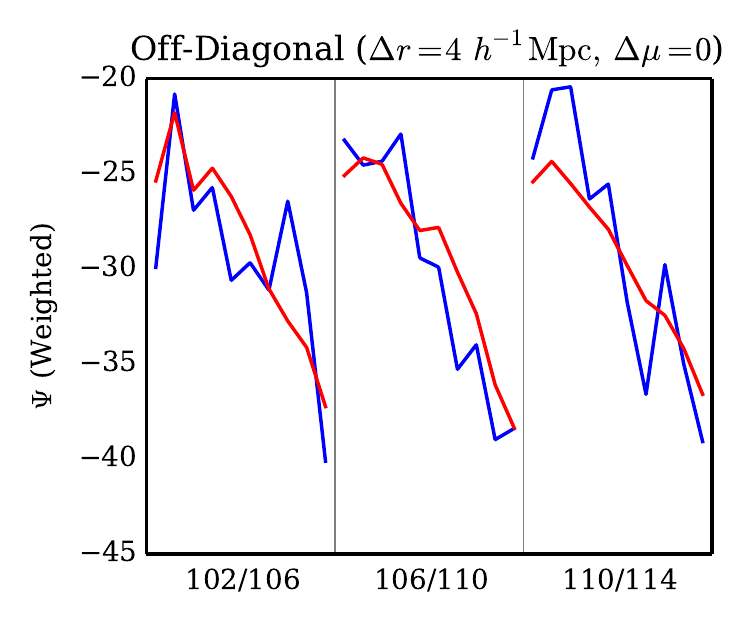}

\caption{\label{fig:PrecLines}We plot here the most prominent parts of the
precision matrices in fig. \ref{fig:PrecBoxes}, both from the mocks
(blue) and from our non-Gaussian model (red, fit using the $\mathcal{L}_{2}$
likelihood). }
\end{figure}

\subsection{\label{sub:Results}Results}

After using both likelihoods to fit our model to the mock catalogs,
we find
\begin{eqnarray}
a_{1} & = & 1.145\pm0.002\,,\\
a_{2} & = & 1.131\pm0.005,
\end{eqnarray}
where $a_{1}$ is the shot noise rescaling determined using $\mathcal{L}_{1}$
and $a_{2}$ is determined using $\mathcal{L}_{2}$. We emphasize
that this rescaling should not be interpreted as a mismatch between
the number density in our code and the actual number density in the
mocks, but rather as the additional shot noise required to mimic the
effects of short-scale non-Gaussianity for the covariance matrix.
The error bars for $a_{1}$ and $a_{2}$ are determined by jackknife
resampling on the mocks, and establish that given 1,000 mocks, the
$\mathcal{L}_{1}$ and $\mathcal{L}_{2}$ likelihoods do lead to different
rescalings of the shot noise.

One way to test the goodness-of-fit of our models is to determine
what distribution of the KL divergence would result from noisy realizations
of our best-fitting models. In order to ensure that the distributions
we generate are comparable to the mocks, we do the following:
\begin{enumerate}
\item Draw 1,000 correlation functions that follow our best-fitting covariance
matrix.
\item Compute the sample covariance, $C_{\mathrm{noisy}}$, and corresponding
precision matrix for those draws.
\item Fit $C_{\mathrm{noisy}}$ or $\Psi_{\mathrm{noisy}}$ as appropriate,
to generate $C_{\mathrm{refit}}$ and $\Psi_{\mathrm{refit}}$
\item Compute the KL divergence between the $C_{\mathrm{noisy}}/\Psi_{\mathrm{noisy}}$
and $\Psi_{\mathrm{refit}}/C_{\mathrm{refit}}.$
\end{enumerate}
The results of these tests are collected in table \ref{tab:KL}. We
see that fitting with either likelihood provides a significant improvement
over the purely Gaussian model. As anticipated, we find that fitting
with the two likelihoods leads to small discrepancies in the resulting
KL divergences. We also find that after fitting, the differences between
the mock and model covariance matrices are too large to be consistent
with noise, indicating that there is room for further improvement
in our models.
\begin{table}
\begin{centering}
\begin{tabular}{|l|c|c|}
\hline 
 & $C_{\mathrm{\ensuremath{test}}}=C_{\mathrm{mock}}$ & $C_{\mathrm{test}}=C_{\mathrm{noisy}}$\tabularnewline
\hline 
$KL\left(\Psi_{\mathrm{G}},C_{\mathrm{test}}\right)$ & 42.14 & 35.13$\pm$0.17\tabularnewline
\hline 
$KL\left(\Psi_{\mathrm{NG}}\left(a_{1}\right),C_{\mathrm{test}}\right)$ & 37.66 & 35.13$\pm$0.17\tabularnewline
\hline 
$KL\left(\Psi_{\mathrm{NG}}\left(a_{2}\right),C_{\mathrm{test}}\right)$ & 37.69 & 35.13$\pm$0.17\tabularnewline
\hline 
$KL\left(\Psi_{\mathrm{test}},C_{\mathrm{G}}\right)$ & 45.76 & 40.59$\pm$0.27\tabularnewline
\hline 
$KL\left(\Psi_{\mathrm{test}},C_{\mathrm{NG}}\left(a_{1}\right)\right)$ & 42.70 & 40.59$\pm$0.27\tabularnewline
\hline 
$KL\left(\Psi_{\mathrm{test}},C_{\mathrm{NG}}\left(a_{2}\right)\right)$ & 42.66 & 40.59$\pm$0.27\tabularnewline
\hline 
\end{tabular}
\par\end{centering}

\caption{\label{tab:KL}Kullback-Leibler (KL) divergences between our models
(Gaussian, and Non-Gaussian fit using $\mathcal{L}_{1}$ and $\mathcal{L}_{2}$)
and the mock catalogs. In order to calibrate, we have also computed
KL divergences between each model and fits to noisy realizations of
that model. With 1,000 mocks it is clear that the mock covariance
matrix is \emph{not} consistent with noisy realizations of our models.}
\end{table}

In order to determine whether our model is sufficient to be used in
studies of BAO, we return to the Fisher matrix approach introduced
in sec. \ref{sub:Comparison-with-Mocks}. After computing the error
ellipses, we would like to know whether the error ellipse from the
mocks is consistent with a noisy draw from the model covariance matrix.
To determine this we follow a procedure similar to the one described
above: we generate sets of 1,000 mock correlation functions drawn
following the model covariance matrix, compute the sample covariance
for those noisy draws, and then compute the error ellipse from that
sample covariance. By repeating this procedure, we can arrive at uncertainties
in the uncertainties in $\alpha$ and $\epsilon$, $\sigma_{\alpha}$
and $\sigma_{\epsilon}$, as well as the uncertainty in their correlation
coefficient, $r_{\alpha\epsilon}$. These results are tabulated in
table \ref{tab:Ellipses}. We find that when focusing only on the
$\alpha-\epsilon$ error ellipses, and with only 1,000 mocks, the
models fit using $\mathcal{L}_{1}$ and $\mathcal{L}_{2}$ are indistinguishable,
and are consistent at the $1\sigma$ level with the QPM mocks. This
constitutes an improvement over the unfit Gaussian model, which is
$>2\sigma$ discrepant from the QPM mocks. Thus, while we would be
reluctant to use the Gaussian model in some applications, the simple
non-Gaussian model introduced here is suitable for use in contemporary
BAO studies. 
\begin{table}
\begin{centering}
\begin{tabular}{|l|c|c|c|c|c|c|}
\hline 
 & $\sigma_{\alpha}$ & $\sigma_{\epsilon}$ & $r_{\alpha\epsilon}$ & $\sigma_{\alpha}$ (noisy) & $\sigma_{\epsilon}$ (noisy) & $r_{\alpha\epsilon}$ (noisy)\tabularnewline
\hline 
\hline 
$C_{\mathrm{mock}}$ & -- & -- & -- & 0.0403 & 0.0422 & 0.189\tabularnewline
\hline 
$C_{\mathrm{G}}$ & 0.0384 & 0.0391 & 0.216 & 0.0387$\pm$0.0011 & 0.0394$\pm$0.0011 & 0.217$\pm$0.038\tabularnewline
\hline 
$C_{\mathrm{NG}}\left(a_{1}\right)$ & 0.0400 & 0.0405 & 0.203 & 0.0403$\pm$0.0011 & 0.0409$\pm$0.0012 & 0.204$\pm$0.038\tabularnewline
\hline 
$C_{\mathrm{NG}}\left(a_{2}\right)$ & 0.0399 & 0.0404 & 0.204 & 0.0401$\pm$0.0011 & 0.0407$\pm$0.0011 & 0.205$\pm$0.038\tabularnewline
\hline 
\end{tabular}
\par\end{centering}

\caption{\label{tab:Ellipses}Fisher matrix results for the uncertainties $\sigma_{\mbox{\ensuremath{\alpha}}}$
and $\sigma_{\epsilon}$, as well as the correlation coefficient $r_{\alpha\epsilon}$.
Even the Gaussian (unfit) model produces reasonable agreement in these
parameters, and the level of agreement improves for the non-Gaussian
(fit) models. When we restrict our attention to the BAO parameters,
we find that the mock precision matrix \emph{is} consistent with a
noisy realization of our fit models, at least when we are limited
to 1,000 mocks.}
\end{table}

\section{Outlook}

We have demonstrated that integrating a simple Gaussian model for
the covariance matrix of the galaxy correlation function yields a
result that is in surprising agreement with that obtained from sample
statistics applied to mock catalogs. A simple extension of that model,
wherein the shot noise is allowed to vary and calibrated against mock
catalogs, further improves that agreement, so that for the purposes
of BAO measurements the mock and model covariance matrices are statistically
consistent with each other. This was accomplished with an effective
number of mocks $N_{\mathrm{eff}}\sim35,000$ using just $\approx1,000$
CPU hours, rather than the hundreds of thousands of CPU hours required
to generate and analyse 1,000 QPM mocks. Based on these results, we
consider this method of covariance matrix estimation to be a strong
alternative to producing very large numbers of mock catalogs. While
$\mathcal{O}\left(100\right)$ mocks will continue to be necessary
for estimating systematic errors, we believe those $\mathcal{O}\left(100\right)$
mocks will be sufficient to calibrate our model, or successors to
it. With a reduction in the \emph{number} of mocks required, we anticipate
greater effort (both human and computational) can be devoted to improving
the \emph{accuracy} of the mocks, ultimately improving the reliability
of error estimates from future analyses.

For the sake of simplicity we have focused on reproducing the mock
covariance matrix. This is a somewhat artificial goal, since the mock
catalogues are themselves imperfect representations of an actual survey.
Turning our attention to the survey covariance matrix, we find an
opportunity and a challenge. In our method, the mocks are only used
to calibrate non-Gaussian corrections, in the form of increased shot
noise, to the covariance matrix. Since we expect the effects of non-Gaussianity
to be confined to relatively short scales, we can imagine performing
this calibration step against simulations with a smaller volume and
higher resolution, compared to the mocks. At the same time, we face
the challenging question of how to bound possible discrepancies between
our model covariance matrix and the true survey covariance matrix.
While the accuracy requirements for a covariance matrix (whether derived
from mocks or from a model) are ultimately determined by the science
goals of a survey, the question of how to determine those requirements
and assess whether they have been reached seems largely open.

A clear benefit of mock catalogs is that the same set of mocks can
be used for measurement using a variety of observables (e.g. the power
spectrum) and cosmological parameters (e.g. $f\sigma_{8}$, often
measured using redshift space distortions). While we have constructed
a model covariance matrix suitable for analysis of the galaxy correlation
function at BAO scales, our goal of reducing the number of mocks required
for future surveys cannot be realized until analogous methods are
developed for the other analyses typical of modern surveys. The covariance
matrix of the power spectrum has received theoretical attention from
many authors \citep{Takahashi2009,Takahashi2011,de-Putter:2012aa,Mohammed:2014lja}.
Recent efforts \citep{Carron:2014hja} have come close to presenting
a usable model for the covariance matrix of the power spectrum, though
the effects of a non-uniform window function have not yet been incorporated.
We are optimistic that the necessary extensions of those models can
be performed, reducing the required number of mocks for power spectrum
analyses to be in line with the $\mathcal{O}\left(100\right)$ necessary
for correlation function analyses.

\section*{Acknowledgements}

RCO is supported by a grant from the Templeton Foundation. DJE is supported by grant DE-SC0013718 from the U.S. Department of Energy. SH is supported by NASA NNH12ZDA001N- EUCLID and NSF AST1412966. NP is supported in part by DOE DE-SC0008080.

Funding for SDSS-III has been provided by the Alfred P. Sloan Foundation, the Participating Institutions, the National Science Foundation, and the U.S. Department of Energy Office of Science. The SDSS-III web site is http://www.sdss3.org/.
SDSS-III is managed by the Astrophysical Research Consortium for the Participating Institutions of the SDSS-III Collaboration including the University of Arizona, the Brazilian Participation Group, Brookhaven National Laboratory, Carnegie Mellon University, University of Florida, the French Participation Group, the German Participation Group, Harvard University, the Instituto de Astrofisica de Canarias, the Michigan State/Notre Dame/JINA Participation Group, Johns Hopkins University, Lawrence Berkeley National Laboratory, Max Planck Institute for Astrophysics, Max Planck Institute for Extraterrestrial Physics, New Mexico State University, New York University, Ohio State University, Pennsylvania State University, University of Portsmouth, Princeton University, the Spanish Participation Group, University of Tokyo, University of Utah, Vanderbilt University, University of Virginia, University of Washington, and Yale University.

This research used resources of the National Energy Research Scientific Computing Center, a DOE Office of Science User Facility supported by the Office of Science of the U.S. Department of Energy under Contract No. DE-AC02-05CH11231.

\bibliographystyle{mnras}
\bibliography{CovRefs}

\end{document}